# Machine learning traction force maps of cell monolayers

Changhao Li[1], Luyi Feng[1], Yang Jeong Park[2,3], Jian Yang[4], Ju Li[2], and Sulin Zhang[1,4,5*]

[1]Department of Engineering Science and Mechanics, Pennsylvania State University, University Park, PA, USA.

[2]Department of Nuclear Science and Engineering, and Department of Materials Science and Engineering, Massachusetts Institute of Technology, Cambridge, MA 02139, USA

[3]Department of Electrical and Computer Engineering, and Institute of New Media and Communications, Seoul National University, Gwanak-gu, Seoul, 08826, Republic of Korea

[4]Department of Biomedical Engineering, Pennsylvania State University, University Park, PA, USA.

[5]Department of Material Science and Engineering, Pennsylvania State University, University Park, PA, USA.

**Abstract:**

Cellular force transmission across a hierarchy of molecular switchers is central to mechanobiological responses. However, current cellular force microscopies suffer from low throughput and resolution. Here we introduce and train a generative adversarial network (GAN) to paint out traction force maps of cell monolayers with high fidelity to the experimental traction force microscopy (TFM). The GAN analyzes traction force maps as an image-to-image translation problem, where its generative and discriminative neural networks are simultaneously cross-trained by hybrid experimental and numerical datasets. In addition to capturing the colony-size and substrate-stiffness dependent traction force maps, the trained GAN predicts asymmetric traction force patterns for multicellular monolayers seeding on substrates with stiffness gradient, implicating collective durotaxis. Further, the neural network can extract experimentally inaccessible, the hidden relationship between substrate stiffness and cell contractility, which underlies cellular mechanotransduction. Trained solely on datasets for epithelial cells, the GAN can be extrapolated to other contractile cell types using only a single scaling factor. The digital

TFM serves as a high-throughput tool for mapping out cellular forces of cell monolayers and paves the way toward data-driven discoveries in cell mechanobiology.

**Main Text:**

The remarkable ability of cells to sense, respond, and adapt to mechanical forces is essential to cellular functions[1, 2, 3, 4]. Much has been known about mechanical force transduction that entangles with biochemical signaling in cell-cell and cell-extracellular matrix communications, directing long-range multicellular morphogenesis[5, 6, 7], wound healing[8, 9], and cancer metastasis[10, 11]. However, the significance of mechanical force transduction in cellular mechanobiology is underserved by the lack of reliable, high-throughput tools for cellular force measurements[12, 13, 14]. Traction force microscopy (TFM) has been widely used to measure the pulling force by focal adhesion points[15, 16, 17]. The obtained traction force map is a prerequisite for monolayer stress microscopy (MSM)[18, 19] and intercellular tension microscopy (ITM)[20, 21, 22]. However, TFM is of low throughput and labor intensive[13, 23] and suffers from decreasing-to-vanishing resolution as the stiffness of the extracellular matrix (ECM) goes above 50 kPa[11, 20, 24]. The lack of a reliable, high-throughput tool for cellular force measurements has critically hindered our understanding of cellular mechanobiology.

Machine-learning (ML) methods have revolutionized pattern recognition and generation with unprecedented accuracy and efficiency[25]. Here, we develop a generative adversarial network (GAN)[26] to learn and predict the traction force distribution of cell monolayers seeding on a flat surface. Taking cell phase contrast images as the input, the GAN analyzes traction force maps as an image-to-image translation problem, where its generative and discriminative neural networks are cross-trained toward a converged solution. As experimental TFM yields insufficient experimental data required for neural network training and testing, we here resort to a recently developed high-throughput continuum model to generate numerical data[20]. Upon optimization and training with the hybrid experimental and numerical datasets, our neural network can perform diverse tasks, including forward prediction of the substrate-stiffness and colony-size dependent traction force maps and inverse delineation of the hidden relationship between cell contractility and substrate stiffness. Furthermore, our ML model predicts traction force concentration at the stiff side when seeding cells on a substrate with stiffness gradient, which implies collective durotaxis. While the GAN is trained by datasets exclusively for epithelial cells, it can be conveniently extrapolated to other contractile cell types by only a single scaling factor. Compared with conventional TFM and physics-based modeling, our digital TFM (DTFM) is highly efficient and versatile to be further adapted to other image-image translation tasks in cell mechanobiology.

**Traction force generation and data preparation**

When seeded on a substrate, migrating cells develop focal adhesion points that pull and deform the substrate. The deformation depends on the pulling force of the cells as well as the mechanical properties of the substrate, i.e., its stiffness and Poisson's ratio[11]. For two-dimensional (2D) TFM that measures the tangential pulling force, i.e., traction on flat substrates[15, 16], fluorescent beads are embedded into a soft substrate (e.g., hydrogels) to track the traction-induced displacement through an optical microscope,[15, 27] in reference to the traction-free condition in which cells are detached off from the substrate. The measured displacement field along with the boundary conditions furnishes an inverse elasticity problem for traction force calculation[11, 24]. Using the TFM-reconstructed traction force distribution as a force boundary condition, the stress in the cell body, modeled as a thin monolayer,[19, 20, 28] can be further determined by monolayer stress microscopy (MSM). The traction force profile indicates focal adhesion distribution[16, 29], while the direction of the first principal stress of the cell monolayer infers stress-fiber orientation[19, 30]. For migrating cells, TFM can be applied at different time points, thereby obtaining time-varying traction force maps[31, 32].

Training an ML model on high dimensional spaces entails a large dataset for converged model performance. For each cell type, we have no more than 100 experimental traction force maps available up to date, with epithelial cells being the mostly characterized cell type (**Extended data Fig. 1**). Preparing a large database ($10^3$~$10^4$ traction force maps) through experimental TFM is undoubtedly very costly, owing to its low-throughput characteristic. To meet this challenge, we here generate training data by invoking our previously developed continuum model[20]. The continuum model assumes that the traction force distribution of a cell monolayer is determined by its current configuration[33], and imposes the chemomechanical balance laws on the cell-substrate system (Supplementary Information). The continuum model simultaneously yields focal adhesion distribution, monolayer stress distribution, and traction force maps. The predicted traction force maps and monolayer stress distributions (**Extended data Fig. 2**) agree very well with TFM and MSM measurements, respectively.[20]

Based on the continuum model, we built a high-throughput simulation workflow to obtain traction force maps with randomly generated cell profiles (Methods, Supplementary Information). The numerical results constitute the main part of the dataset, in addition to the experimental traction force maps from our own group and others. To complement the dataset, we also performed the standard data augmentation algorithm (Supplementary Information) on the limited experimental data. The final dataset includes ~8000 numerical simulations and ~600 experimental TFM measurements.

**Machine-learning algorithms**

Here we regard traction force as a configurational force[33], determined by cell geometry, substrate stiffness, and cell type, which constitute the training data of our ML model. The input cell boundary $r(\theta)$ can be regularized as a $n \times n$ binary geometric indicator tensor $\boldsymbol{\Omega}$ where the element values take 1 inside the cell boundary or 0 otherwise, and $n$ varies for different resolutions. The size of the cell colony, represented by its average radius $\bar{R}$, is given by $\bar{R} = \sqrt{S_{\text{cell}}/\pi}$, where $S_{\text{cell}}$ is the area occupied by the cell monolayer. Correspondingly, the traction force vector field $\boldsymbol{T}(\boldsymbol{x})$ can be discretized as a $n \times n \times 2$ tensor, where the two channels in the third dimension represent the traction forces along the in-plane orthogonal directions ($x$ and $y$ directions). Other scalar-valued parameters, such as substrate stiffness, cell radius, and cell contractility, are combined into the aggregated property tensor $\boldsymbol{C}$.

Combining experimental measurements and continuum mechanics modeling, we sample the input space of $\boldsymbol{\Omega}$ and $\boldsymbol{C}$. In experiments, $\boldsymbol{T}$ is directly measured by TFM. In modeling, we use high-throughput simulations to obtain the traction force field $\boldsymbol{T}$ (**Fig. 1A**, Supplementary Information). Based on the hybrid database, we propose two ML tasks: I) Given cell geometry $\boldsymbol{\Omega}$ and the parameters $\boldsymbol{C}$, predict the traction force map $\boldsymbol{T} = f(\boldsymbol{\Omega}, \boldsymbol{C})$; II) Given $\boldsymbol{T}$ and $\boldsymbol{\Omega}$, predict a set of parameters $\boldsymbol{C} = \hat{f}(\boldsymbol{\Omega}, \boldsymbol{T})$, some of which are hidden relationships that are inaccessible to experiments and underlie cellular mechanotransduction.

For task I, we recast the prediction of traction force maps from given cell geometry as the image-image translation problem[34] in computer vision technology. Inspired by the recent progress in image generation[35] and style transfer[36], we designed a generative adversarial network (GAN) to extract the underlying distribution patterns of traction force. A GAN consists of a generative (generator) and a discriminative (discriminator) neural network, where the generator and the discriminator compete and evolve together to reach a final converged image. Upon training, the generator gains increasing ability to generate artificial data that can effectively fool the discriminator. When the discriminator can no longer effectively distinguish the difference between the ground-truth data and artificial data generated by the generator, the cross-training converges. As shown in **Fig. 1B-C**, our generator is adapted from U-Net[37], a classical convolutional neural network (CNN) widely applied in biomedical image segmentation tasks, and our discriminator is composed of a CNN linked to a fully connected neural network. The loss function $L(D,G)$ of the GAN can be formulated by the cross entropy,

$$L(D,G) = \mathbb{E}_{T \sim p(T)}[\log(D(T))] + \mathbb{E}_{\Omega,C \sim p(\Omega,C)}\left[\log\left(1 - D(G(\Omega,C))\right)\right] \qquad (1)$$

where $D$ and $G$ represent the discriminator and generator, respectively, and $\mathbb{E}$ denotes the pixelwise average over the training set. **Fig. 1C** shows the workflow and dataflow of the proposed GAN, where the gradients of $L(D,G)$ backpropagate to the discriminator $D$ and the generator $G$, enabling their simultaneous training process: $\max_D \min_G L(D,G)$. The discriminator outputs a binary reliability matrix indicating the extent of "validity" of local segments of the input images **(Fig. 1B)**, while the generator $G$ predicts the traction force map $T$ from the given cell geometry $\Omega$ and the property tensor $C$.

**Machine-learning prediction of traction force maps**

As shown in **Fig. 1**, we train the GAN on the hybrid database (Methods and Supplementary Information), which includes simulated and experimentally measured traction force maps for different cell colony sizes and substrate stiffnesses. Our training data are solely based on epithelial cells, as this cell type has the most available datasets of experimental traction force maps. The traction force map predicted by GAN shows great consistency with both simulation and TFM results for different colony sizes (**Fig. 2A**) and substrate stiffnesses (**Fig. 2B**). The overall profile of predicted traction force magnitude follows the similar trend of exponential decay from the cell periphery to its center and accumulates in the boundary regions with higher local convex curvature (**Fig. 2A-B**, **Extended data Fig. 3**). Above two features of traction force distribution can also be clearly observed from the intermediate feature maps (**Extended data Fig. 4**) from the trained generator. We then define the average pixelwise error by $\epsilon_T = \frac{1}{n^2 N} \Sigma_N \left( \left| \left( T^p_{ijk} - T^t_{ijk} \right) \right| / \left| T^t_{ijk} \right| \right)$, where $T_{ijk}$ denotes the components of the ML-predicted ($T^p_{ijk}$) or the ground-truth ($T^t_{ijk}$) traction force tensor, and $N$ is the size of the training set. $\epsilon_T$ was measured separately on the test datasets from the continuum simulation and TFM experiments. We found that for all cases $\epsilon_T < 15\%$ and monotonically increases with the colony size $\bar{R}$ (**Fig. 2C**), but relatively stable with substrate stiffness $E_s$ (**Fig. 2D**). Indeed, given a fixed size $n$ of geometry indicator $\Omega$, a larger $\bar{R}$ corresponds to downsampling and hence decreases resolution. As expected, the GAN achieves better performance on the simulation test set, which constitutes the main part of the training data.

**Predicting collective durotaxis**

In durotaxis, migrating cells sense and follow environmental stiffness gradient[38, 39], exhibiting an asymmetrically localized traction force distribution at the leading and trailing edges that signifies cell migration direction[38, 40]. The asymmetric traction force distribution can be attributed to the maturation of focal adhesion points from which lamellipodia extend forward for cell crawling[41, 42]. Here we explore if our ML model can predict the traction force localization $T_{\text{pred}}(x) =$ $f(E_{\mathrm{s}}(\pmb{x}), \pmb{\Omega})$ with given substrate gradient $\nabla E_{\mathrm{s}}(\pmb{x})$ and cell geometry $\pmb{\Omega}$ (**Fig. 3A**). We sample the input space of $E_{\mathrm{s}}(\pmb{x})$ by 5000 additional simulations and train the GAN on the extended dataset (Supplementary Information). **Fig. 3B** shows that our ML model captures the asymmetric traction force distribution of the cell monolayer on a substrate with a stiffness gradient. Specifically, traction force localization occurs more predominantly on the stiff side of the substrate, indicating soft-to-rigid cell migration direction, i.e., durotaxis, consistent with the reported experimental observations[38, 40, 43].

It was previously revealed that collective durotaxis arises from long-range transmission of intercellular forces, where a multicellular colony can sense weak stiffness gradient of the substrate but isolated individual cells cannot[38, 43]. To see whether our ML model can predict collective durotaxis, here we vary substrate stiffness gradient **(Fig. 3C)** and colony size **(Fig. 3D)** and quantify the traction force difference $\Delta \pmb{T} = \pmb{T}_{\mathrm{G}} - \pmb{T}_0$, where $\pmb{T}_{\mathrm{G}}$ and $\pmb{T}_0$ are the predicted traction force maps by our ML model for the same cell colony on the substrate with stiffness gradient and uniform substrate with the same average stiffness, respectively. In **Fig. 3C-D**, we visualize the horizontal component of $\Delta \pmb{T}$, $\Delta T_{\mathrm{x}}$, the same direction as the stiffness gradient. Our ML model predicts pronounced localization of traction forces at the stiff side with increasing substrate stiffness gradient and cell colony size. As lamellipodia extend from the localization sites of traction force, our prediction indicates collective durotaxis, consistent with previously reported results[38].

**Extracting exogenous cell contractility and substrate stiffness**

Cells sense and adapt to their mechanical environments by operating their contractile machinery at different levels through mechanotransduction[6, 44, 45]. Thus, cell contractility and substrate stiffness are intimately correlated[46, 47], though it remains a challenge to quantify such a relationship experimentally. Here we recast our ML task to (II), inversely predicting hidden correlations using traction force map $T: C = \hat{f}(\Omega, T)$. In the inverse prediction, traction force maps are taken as known data, but $E_s(x)$ and the cell contractility $\sigma_A$ are variables to be learned (**Fig. 4A**). Similarly, by resorting to the continuum simulations, we sample the input spaces of $T$, $\Omega$ and the corresponding output $\sigma_A$ and $E_s(x)$, and train a U-Net on this dataset (Supplementary Information). **Fig. 4B** shows full-field stiffness prediction for circular and linear patterning modes of substrate stiffness. Our machine learning prediction for $E_s(x)$ shows qualitative consistency for the whole range of substrate stiffnesses, and successfully captures the spatial variations of substrate stiffness. **Fig. 4C** displays the comparison between the ground truth of the substrate modulus $\bar{E}_s$ used in continuum simulations and the predicted $\bar{E}_{s,\text{pred}}$. Furthermore, we find that the U-Net can learn the underlying nonlinear function between active stress $\sigma_A$ and the substrate stiffness $E_s$ (**Fig. 4D**). This inverse procedure provides a new route to estimate spatially varying properties of the cell-substrate system.

**Extrapolating the ML model to other cell types**

Contractile cells of different types share similar mechano-biochemical feedback loops in traction force generation[48, 49], despite cell-type dependent contractility $\sigma_A$ is in response to the local mechanical environment such as substrate stiffness. Thus, we hypothesize that traction force maps for all the types of contractile cells have similar spatial patterns, but scale with cell-type dependent $\sigma_A$. This provides a convenient route to further extend our ML model to other cell types. Without any retraining, we directly scale the trained GAN to four different cell types (fibroblast[50], Hela[51], MDCK[22], and Osteosarcoma[52] cell lines) by a constant α and compare the prediction with the reported experimental measurements (**Fig. 5A-B**). As expected, the predicted traction force maps of the GAN, which is only trained on HCT-8 epithelial datasets, have reasonably good agreement with the experimental measurements, demonstrating the transferrable learning abilities of our ML model.

**Discussion**

In summary, the GAN, trained by hybrid experimental and numerical datasets, can paint out high-fidelity traction force maps using only cell contours as the input. As the continuum model involves multi-field coupled partial differential equations (PDEs), our ML framework serves as an efficient PDE solver. In addition to predicting substrate-stiffness and colony-size dependent traction force maps, the neural network can unveil the hidden relationship between substrate stiffness and cell contractility, which is experimentally inaccessible but at the heart of mechanobiology. Without further modification of neural network architecture, the ML model predicts asymmetric traction force distribution in accordance with substrate stiffness gradient, demonstrating its power of predicting collective durotaxis. Furthermore, trained by data only for epithelial cells, our ML model can be extended to other contractile cell types, by only a single scaling factor associated with each cell type. Taken together, the GAN presents a powerful DTFM for estimating the traction force maps of cell monolayers.

Our DTFM can be employed to predict the time-varying traction forces for migrating cells. This extension is rather straightforward as cell contour is the only required input for the DTFM and is routinely available in almost all cell biology labs. As traction force distribution is a prerequisite for measurements of cell monolayer stress and intercellular tension, the established DTFM enables convenient development of digital MSM and IFM, making it a complete toolset for extra-, intra- and inter-cellular force measurements. Such digital force microscopies support new discoveries with unprecedented pace and precision with regard to the central role of mechanical forces in complex mechanobiochemical processes, such as collective cell migration[38, 53, 54], multicellular morphogenesis[5, 55], and cancer metastasis[10, 11, 42], etc.

**Methods**

**Design of computational experiments**

To complement the TFM measurements, we construct a database of cell profiles and corresponding traction forces using continuum simulations (Supplementary Information). To generate a sufficiently extensive dataset that can cover the latent space of 2D cell profiles, we develop an algorithm to effectively control the regularity and curvature of randomly generated cell geometries. The algorithm uses a simple descriptor vector $\boldsymbol{d} = (n_s, \bar{R}, r_s, \sigma_s)$, where $n_s$ refers to the number of boundary waves, $\bar{R}$ the average radius of the cell, $r_s$ the fluctuation for the radius of curvature, and $\sigma_s$ the Gaussian smoothing parameter. By defining a suitable range and discretized interval for the parameter space embedded in the four-dimensional vector $\boldsymbol{d}$, we generate a cell profile dataset with 7000 samples, which is utilized as the geometry for the continuum model. We develop Python scripts to automate the above algorithm for high-throughput simulations. Further information is available in the Supplementary Information.

**Continuum model of cell traction forces**

Our continuum model for cell traction forces is based on Reference[20], where we consider the chemomechanical thermodynamics equilibrium of a cell monolayer adhering to a soft substrate. Beginning from a Helmholtz free energy functional incorporating the elastic, edge, and interfacial effects of the cell-substrate system, along with the biochemical signaling and diffusion of the adhesion molecules, we analyze the stable state by minimizing the total free energy with respect to the displacement field and the concentration of adhesion molecules. See Supplementary Information for details.

We derive the weak form of governing equations (Supplementary Information) and implement the corresponding high-throughput simulations by Java 11.0 and COMSOL Multiphysics package (Version 5.5). The input cell geometries are discretized by quadradic triangular elements with appropriate size balancing the computational cost and simulation precision. Stationary solver with the Newton-Raphson algorithm is applied to solve the nonlinear boundary value problem. The output data from COMSOL Multiphysics is postprocessed by Python scripts to fit the format requirement of machine learning libraries.

**Traction force microscopy**

We seed HCT-8 cells onto a soft substrate made of PAA hydrogels at a density of 2000 cells/cm$^2$. The stiffness of the substrate is controlled by adjusting the concentrations of acrylamide and bis-acrylamide. Prior to cell seeding, we coat the top surface of the substrate with fibronectin. To track the displacements induced by cell traction, we embed fluorescent beads in a single plane beneath the top surface of the substrate. The seeded HCT-8 cells are then cultivated for 24-72 hours, until the cell monolayer has sufficiently adhered, spread, and grown into multicellular colonies of varying sizes. To measure the monolayer thickness, we take three-dimensional images of the cell colonies using a laser-scanning confocal fluorescence microscope (Olympus FV10i, Japan). Standard TFM is applied then solve the inverse elasticity problem for the cell traction forces. Further details on this experimental setup can be found in Supplementary Information.

**Architecture of neural networks**

We design the generator as an adapted U-Net[37], which is an encoder-decoder architecture with a symmetric input and output format. The basic building blocks encoder and decoder of the generator consist of several convolution/max pooling operators and a nonlinear ReLU activation function. The discriminator shares a similar structure to the encoder of the generator. The GAN is implemented based on PyTorch[56], a Python-based deep learning library. We set the training parameter as follows: 200 epochs to stop training, ADAM optimizer with the learning rate $\epsilon = 0.0002$, hyperparameters $\beta_1 = 0.6$ and $\beta_2 = 0.95$. The entire training process is deployed on Google Colab.

**Acknowledgments**: Research reported in this publication was supported by National Institute of Neurological Disorders and Stroke (NINDS) Award NS123433.

**Author contributions:** S.Z. conceptualized the project. C.L. and L.F. collected the experimental data. C.L. designed the high-throughput simulation and the machine learning workflow. C.L., L.F., Y.J.P, and S.Z. performed the data analysis. All authors contributed to the writing of the manuscript.

**Competing interests:** The authors declare that they have no competing interests.

**Supplementary Materials:** Supplementary information is available for this paper.

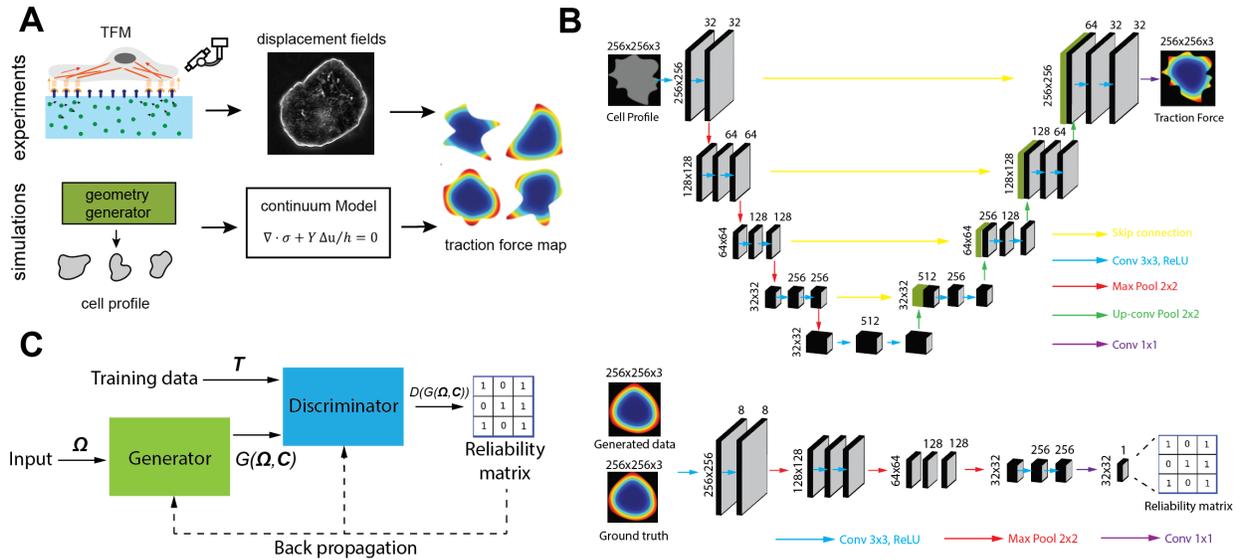

**Fig. 1. Schematic illustration of the data collection processes and the generative adversarial network (GAN). A.** The hybrid data generation processes from TFM experiments and high-throughput simulations. **B.** Top: The architecture of the generator. 3D image tensors are visualized as black cuboids labeled with its size, and neural network operators are represented as arrows with different colors. Yellow arrow: skip connection operator which skips middle network blocks and directly concatenates the input tensor to its output. Blue arrow: $3 \times 3$ convolution operator followed by ReLU activation function, which shrinks the first two dimensions and expands the third dimension of the input tensor. Red arrow: max pooling operator which is similar with convolution operator but takes the local maximum value. Green arrow: up-convolution operator, which is the inverse operator of convolution. Purple arrow: $1 \times 1$ convolution operator, which only shrinks the third dimension and leaves the first two dimensions unchanged. Bottom: The architecture of the discriminator. The information flow and data structure are displayed in a similar way as in the generator. **C.** The training workflow of the GAN. Solid arrows denote forward propagation and dashed arrows denote backpropagation.

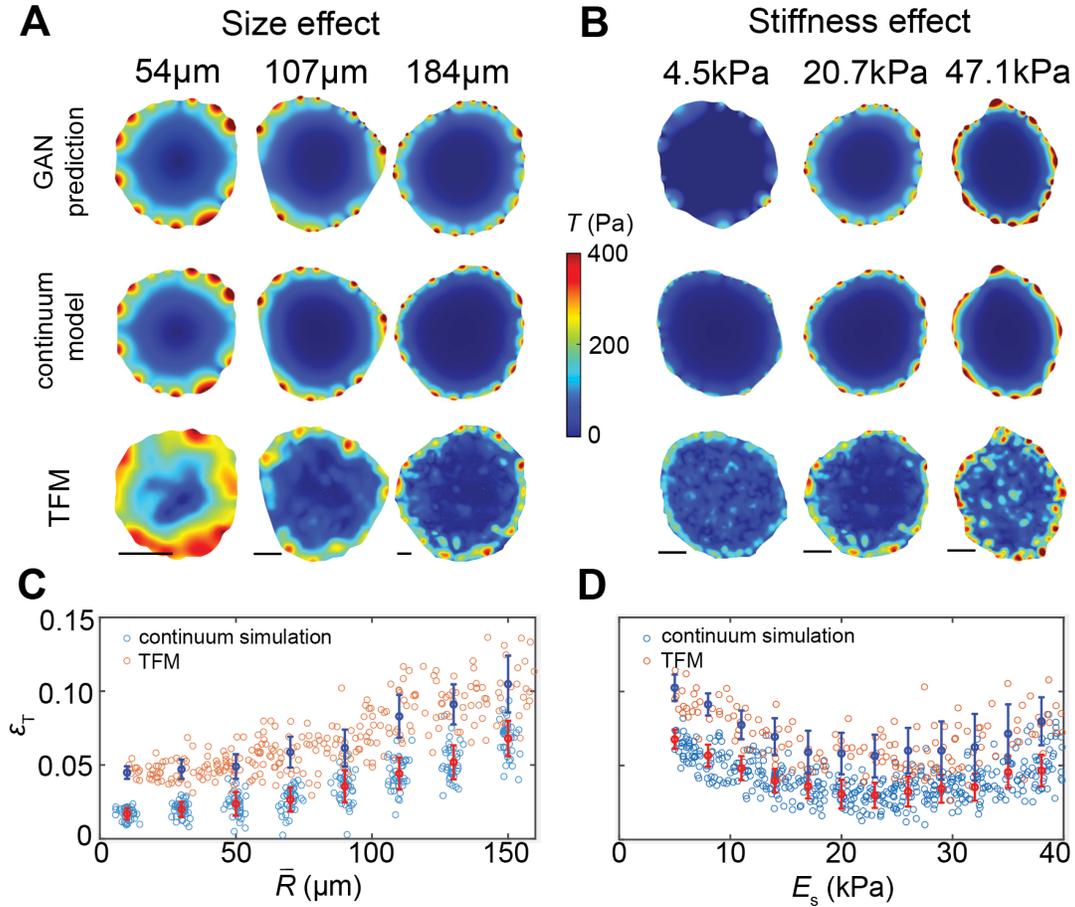

**Fig. 2. ML prediction of cell colony size and substrate stiffness dependent traction force maps. A.** Effect of cell colony size. The traction force maps for cell colonies with different average radius (54 μm, 107 μm, and 184 μm) are experimentally measured (bottom row), simulated (middle row) by the continuum model, and predicted by the ML model (top row). Scale bar: 20 μm. **B.** Effect of substrate stiffness. Colormaps are experimentally measured, simulated, and ML predicted traction force maps of cell colonies with different substrate stiffness (4.5kPa, 20.7kPa, and 47.1kPa), respectively. Scale bar: 20 μm. **C.** Relative prediction error $\epsilon_T$ of our ML model for different average cell colony sizes $\bar{R}$ in comparison to TFM and continuum modeling. **D.** Relative prediction error $\epsilon_T$ for different substrate stiffnesses. Blue (red) error bar denotes the average value and standard deviation of $\epsilon_T$ from experiments (simulations).

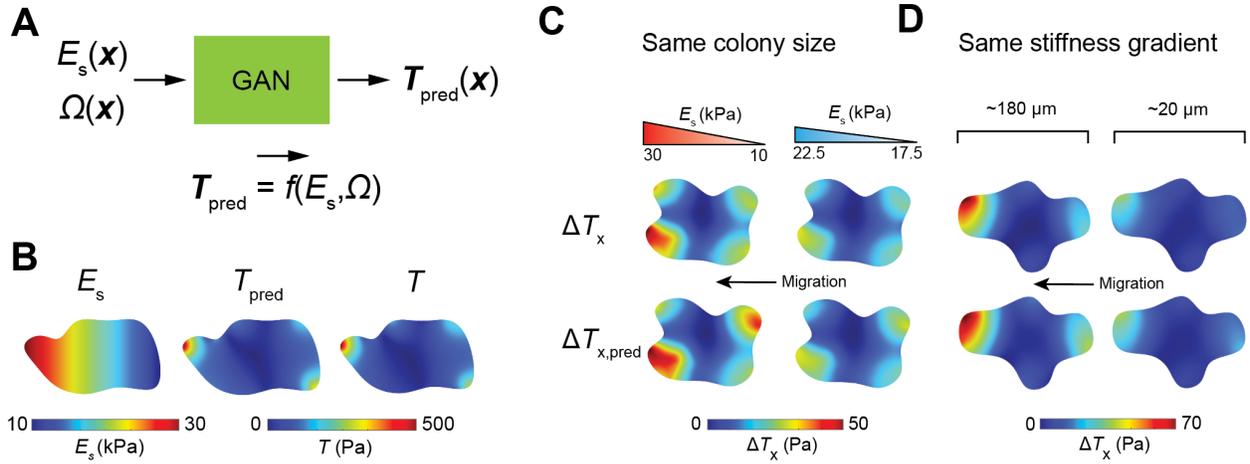

**Fig. 3. The extended ML model predicts collective durotaxis**. **A.** Schematic illustration of forward prediction of traction force maps. **B.** An ML predicted traction force map $T_{\text{pred}}(x)$ from forward prediction with spatially varying substrate stiffness, in good comparison to the continuum simulations. **C-D.** Machine learning predicting collective durotaxis effects. The predicted traction force difference $\Delta T_x(x)$ shows clear localization at the stiff side. The traction force localization becomes more pronounced with increasing stiffness gradient (**C**) and cell colony size (**D**).

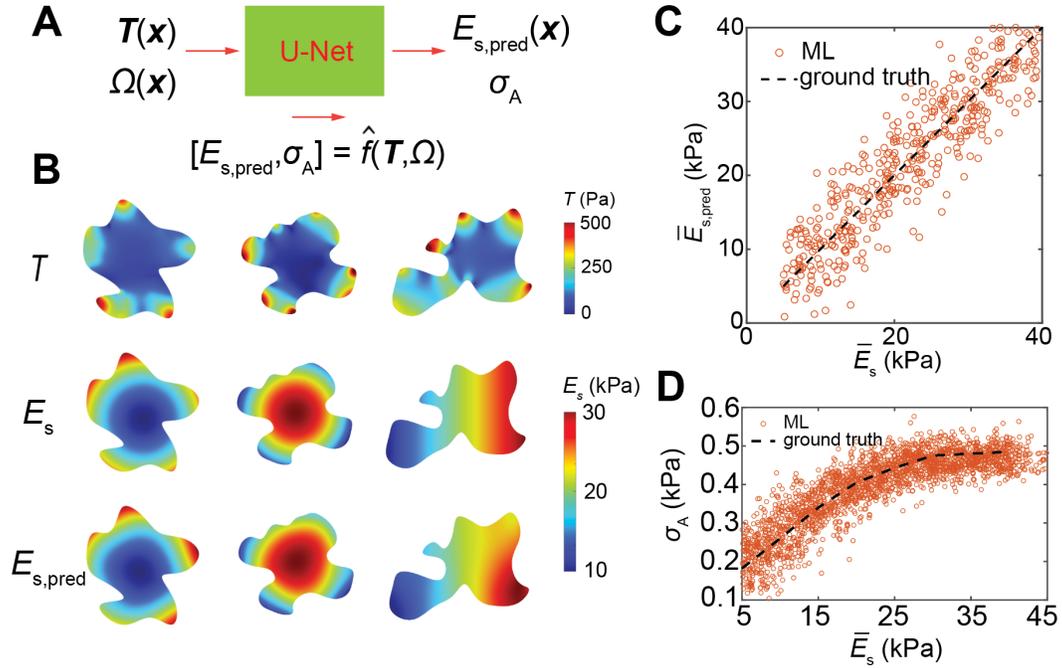

**Fig. 4. ML prediction of the hidden properties of the cell-substrate system**. **A.** Schematic illustration of inverse prediction of cell and substrate properties. **B**. Full-field stiffness prediction results for circular and linear patterning modes of substrate stiffness. **C.** ML regression of spatially averaged stiffness $\bar{E}_s$. The data range for ground truth is computationally generated from 5kPa to 40kPa. **D**. The $\sigma_A - \bar{E}_s$ relation captured by the ML model (hollow scatters), compared with the ground truth generated by simulation (dashed line).

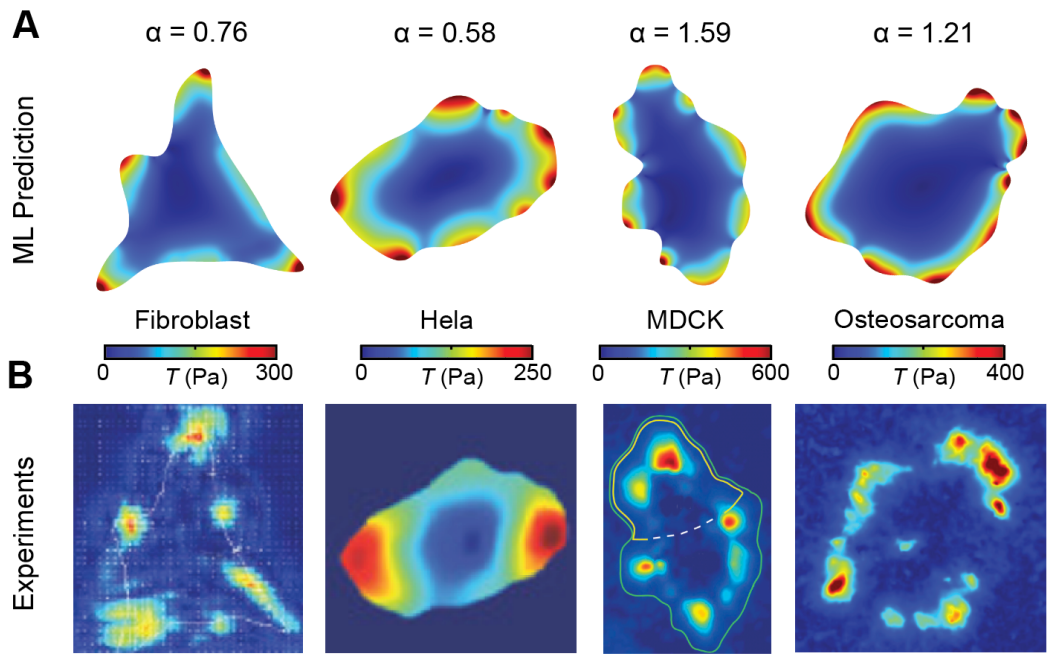

**Fig. 5. Extrapolation of the ML model to other contractile cell types. A-B.** Representative prediction results (**A**) from the GAN, and the comparison (**B**) between experiments. Here $\alpha$ is a scaling factor that accounts for cell-type dependent contractility. Cell geometries and traction force maps of fibroblast[50], Hela[51], MDCK[22] and Osteoarcoma[52] are extracted from previous studies.